\documentclass[11pt,titlepage]{article}

\ifx\pdfoutput\undefined
\usepackage[dvips,bookmarks]{hyperref}  
\else
\usepackage{hyperref}   
\fi

\usepackage[round,numbers,sort&compress]{natbib}

\usepackage[utf8]{inputenc}
\usepackage[english]{babel}

\usepackage{graphicx}
\usepackage{subfigure}
\usepackage{amsfonts}
\usepackage{amssymb,amsfonts,amsmath}
\usepackage{chemarr,color}

\hypersetup{colorlinks,bookmarksopen,bookmarksnumbered,citecolor=blue,
pdfstartview=FitH}

\title{A new mechanism of development and differentiation through slow binding/unbinding of regulatory proteins to the genes}

\author{Haidong Feng$^{a}$,
Jin Wang$^{ab*}$  \\\\
\small{$^a$ Department of Chemistry, Physics and Applied Mathematics} \\
\small{State University of New York at Stony Brook } \\
\small{Stony Brook, NY, 11794, USA } \\
\small{$^b$ State Key Laboratory of Electroanalytical Chemistry } \\
\small{Changchun Institute of Applied Chemistry } \\
\small{Chinese Academy of Sciences } \\
\small{Changchun, Jilin, 130021 } \\
\small{People's Republic of China } \\
\\\\
\small{$^*$ E-mail: jin.wang.1@stonybrook.edu} \\ \\}

\begin{document}

\maketitle
\date{}

\newpage

\begin{abstract}

Understanding the differentiation, a biological process from a multipotent stem or progenitor
state to a mature cell is critically important. We develop a theoretical
framework to quantify the underlying potential landscape and biological paths for cell
development and differentiation. We propose a new mechanism of differentiation and development
through binding/unbinding of regulatory proteins to the gene promoters.
We found indeed the differentiated states can emerge from the slow promoter binding/
unbinding processes. Furthermore, under slow promoter binding/unbinding, we found
 multiple meta-stable differentiated states. This can explain the origin of multiple states observed in the recent
experiments. In addition, the kinetic time quantified by mean first passage transition time for the differentiation
and reprogramming strongly depends on the time scale of the
promoter binding/unbinding processes.
We discovered an optimal speed for differentiation for
certain binding/unbinding rates of regulatory proteins to promoters.
More experiments in the future might be able to
tell if cells differentiate at at that optimal speed.
In addition, we quantify kinetic pathways for the
differentiation and reprogramming. We found that they are irreversible.
This captures the non-equilibrium dynamics in multipotent stem or progenitor cells.
Such inherent time-asymmetry as a result
of irreversibility of state transition pathways as shown may provide the origin of time arrow for cell development.

\end{abstract}

\section{Introduction}

During cell differentiation, the cell evolves
from undifferentiated phenotypes in a multipotent stem or progenitor state to
differentiated phenotypes in a mature cell.
In this process, the gene regulatory network, which
governs the progressive changes of
gene expression patterns of the cell, forces the cell to adopt the
 cell type-specific phenotypes. Cells can have states with the higher probability of appearance, which leads to different cell
phenotypes. Different cell phenotypes correspond to different basins of attractions
on the potential landscape \cite{Waddington, Wang2010, Wang2011}. Therefore the differentiation and developmental process of the cell
can be thought as the evolution of the underlying landscape topography from one basin to to another. One grand challenge is to explain how this occurs, what is the underlying mechanism and how to quantify the differentiation and developmental process. Furthermore, the unidirectional developmental
process posses another challenge to explain the origin of time arrow.


In the cell, intrinsic fluctuations are unavoidable due to the
limited number of protein molecules. There have been
increasing numbers of studies on how the gene regulatory networks can be
stable and functional under such highly fluctuating environments
\cite{Elowitz, WangJCP2007, Arkin}. In addition, the gene state fluctuations
from the regulatory proteins
binding/unbinding to the promoters can be significant for gene expression
dynamics. Conventionally, it was often assumed that the
binding/unbinding is significantly faster than the synthesis and
degradation (adiabatic limit) \cite{Shea, NatureExperiment}. This
assumption may hold in some prokaryotic cells in certain conditions,
in general there is no guarantee
it is true. In fact, one expects in eukaryotic cells and some
prokaryotic cells, binding/unbinding can be comparable or even
slower than the corresponding synthesis and degradation
(non-adiabatic limit).
This can lead to nontrivial stable states and coherent oscillations
appearing as a result of new time scales introduced due to the
non-adiabaticity \cite{Hornos, Walczak, Daniel, Kepler, Kardar,
Feng2011JPCLett, Feng2011JPCB, Feng2012BJ,Weinberger, Xie1}. Therefore,
the challenge for us is to understand how the biological differentiation and reprogramming can be functional under
both intrinsic fluctuations and non-adiabatic fluctuations.

Previous studies showed that
the change in the self activation regulatory strengths can cause the
differentiation of phenotypes \cite{Wang2010, Wang2011, Arias}. In this article, we used a canonical gene regulatory
circuit module to study cell fate decision and commitment in multipotent stem or progenitor cells \cite{Arias, Wang2010, Wang2011}.
We will study a model of cell developmental
circuit (Fig. \ref{circuit}) \cite{Graf}
 which is composed of a pair of
mutually inhibiting but self activating genes.  This gene regulatory motif has been found in various tissues where a pluri/multipotent
stem cell has to undergo a binary cell fate decision \cite{Zhou, Huang}. For
example, in the multipotent common myeloic progenitor cell
(CMP) facing the binary cell fate decision between the
myeloid and the erythroid fate, the fate determining transcription
factors (TF), PU.1, and GATA1, which promote the myeloid or
the erythroid fates, respectively, form such a gene network circuit. The relative
expression levels A (PU.1) and B (GATA1) of these two reciprocal
TFs can bias the decision toward either lineage \cite{Graf, Huang}.

We found that the change in the time scale of the binding/unbinding of regulatory proteins to the promoters may provide an new
important mechanism for the cell differentiation.
We studied the underlying potential landscapes associated with the differentiation and developmental process and found that
the underlying landscapes developed from un-differentiated multipotent
state to the differentiated states as the binding/unbinding rate decreased to the slow non-adiabatic binding region.
In addition, in the slow non-adiabatic binding region, we predicted the emergence of
multiple meta-stable states in the development of multipotent stem cells and explained the origin
of this observation in the experiments \cite{Arias}. We also
calculated the mean first passage transition time for the differentiation and
reprogramming. We found that the mean first passage transition time strongly
depends on the time scale of the promoter binding/unbinding processes.
There is an optimal speed for differentiation and development with certain promoter binding/unbinding rates.
It will be natural to ask whether the differentiation and development
happens at this optimal speed? Future experimental and bioinformatics studies might be able to give the answer.
We quantified the kinetic pathways for the differentiation and reprogramming.
We found that they are irreversible. This captures the non-equilibrium prosperities for the biological processes of the
underlying gene regulatory networks in multipotent stem or progenitor cells. It may provide the origin of time arrow for development.

\section{Methods and Materials}\label{sec.2}

As shown in Fig. \ref{circuit}, the gene regulatory
circuit that governs binary cell fate decision module consists of mutual regulation of two opposing
fate determining master TF A and B. The module has been
shown to control developmental cell fate decision and commitment
in several instances of multipotent stem or progenitor cells
that faces a binary fate decision, (i.e., GATA1 and PU.1)  \cite{Zhou, Huang}.
A and B are coexpressed in the multipotent undecided cell and
committed to either one of the two alternative lineages is associated
with either one factor dominating over the others, leads to
expression patterns in a mutually exclusive manner \cite{Zhou, Hu}.
Importantly, in many cases the genes A and B also self-activate
(positive autoregulate) themselves (Fig. \ref{circuit}).
Here, the hybrid promoter $\alpha$ can be bound by the
regulatory protein $\beta$ with the binding rate $h_{\alpha\beta}$
and dissociation rate $f_{\alpha \beta}$ (both $h_{\alpha\beta}$ and
$f_{\alpha\beta}$ can depend on protein concentration $n_{\beta}$). The
synthesis of protein $\alpha$ is controlled by the gene state of
promoter $\alpha$. There are
two types of genes, $A$ and $B$, to be translated into proteins $A$
and $B$ respectively. The proteins $A$($B$) can bind to the
promoter of the gene $A$($B$) to activate the synthesis rate of
$A$($B$), which makes a self-activation feedback loop.
The proteins $A$($B$) can bind to the gene $B$($A$) to
repress the synthesis rate of $B$($A$), which makes a
mutual repression loop.
Here, both protein $A$ and protein $B$ bind on promoters
as a dimer with the binding rate $\frac{1}{2} h_{\alpha A} n_A (n_A -1)$
and $\frac{1}{2} h_{\alpha B} n_B (n_B-1)$ respectively.
Therefore, each gene has 4 states with self activator binding or non-binding and with mutual repression from another gene binding or non-binding (assuming we have two different binding sites, one for self activator and one for the other gene).
The whole system has 16 gene states in total.
For simplicity, we neglect the roles of mRNAs by assuming translation processes
are very fast. The model can be expressed by the following
chemical reactions:
\begin{eqnarray} \label{assume02}
 &&  \mathcal{O}_{\alpha}^{11} + 2 A
       \xrightleftharpoons[f_{\alpha A}]{h_{\alpha A}} \mathcal{O}_{\alpha}^{01},
       \quad  \mathcal{O}_{\alpha}^{10} + 2 A
      \xrightleftharpoons[f_{\alpha A}]{h_{\alpha A}} \mathcal{O}_{\alpha}^{00} \\
  && \mathcal{O}_{\alpha}^{11} + 2 B
      \xrightleftharpoons[f_{\alpha B}]{h_{\alpha B}} \mathcal{O}_{\alpha}^{10},
      \quad  \mathcal{O}_{\alpha}^{01} + 2 B
     \xrightleftharpoons[f_{\alpha B}]{h_{\alpha B}} \mathcal{O}_{\alpha}^{00}  \\
  && \mathcal{O}_{A}^{ij}
   \overset{g_{A}^{ij}}{\longrightarrow} A, \quad \mathcal{O}_{B}^{ij}
   \overset{g_{B}^{ij}}{\longrightarrow} B,
  \quad A \overset{k_{A}}{\longrightarrow}  \emptyset
  \quad B \overset{k_{B}}{\longrightarrow}  \emptyset \label{assmue2}
\end{eqnarray}
with $\alpha=A (B)$ for the hybrid promoter of gene $A (B)$.
For the gene state index $ij$ of gene $O_\alpha$, the first index
$i=1(0)$ stands for the activator protein $A$ unbound(bound) on
the promoter $\alpha$; the second index $j=1(0)$ stands for the
repressor protein $R$ unbound(bound) on the promoter $\alpha$.
$g_{A}^{ij}$ ($g_{B}^{ij}$) is the synthesis rate of the protein $A$ ($B$)
when the gene $A$ ($B$) is in state $ij$.
The probability distribution of the microstate is indicated as $P_{ijkl} (n_{A}, n_{B})$ where $n_A$ and $n_B$ are the
concentration of the activator
$A$ and the repressor $B$ respectively. The index $i$($j$) represents the
gene $A$ occupation state by the protein $A$($B$) and the index
$k$($l$) represents the gene $B$ occupation state by the
protein $A$($B$). This results sixteen master equations for the probability distribution which are shown explicitly in Supporting Material ({\bf SM}).

The steady state probability distribution satisfies
$\frac{d P^{(ss)}_{ijkl} (n_{A}, n_{R})}{dt} = 0$
for all $i,j,k,l$. The total probability distribution is
$P^{(ss)} = \sum_{ijkl} P^{(ss)}_{ijkl}$.
The generalized potential function $U$ of the
non-equilibrium network can be quantified as: $U(n_A, n_B) = - \ln P^{(ss)}$.
It maps to the potential landscape, which
gives a quantitative measure of the global stability and function
of the underlying network \cite{WangPNAS2008}.
Above equations are difficult to deal with because each one actually represents an infinite number of equations (n range from $0$ to $\infty$).
A direct way to find the steady state $P^{ss}$ is through kinetic simulations \cite{Gillespie}.
Here, we will use Monte Carlo Simulation to find the stead state distribution of master equations (see Supporting Material ({\bf SM})).

\section{Results}\label{sec.3}

In our calculations, we only consider the A-B symmetric case:
\begin{eqnarray}
&h_{AA} = h_{BB}=h_A,  \quad h_{AB} = h_{BA}=h_R \\
&f_{AA} = f_{BB}=f_A,  \quad f_{AB} = f_{BA}=f_R \\
&k_A = k_B = k
\end{eqnarray}
We define the normalized binding/unbinding rate of the gene states: $\omega_A=f_A/k$,  $\omega_R=f_R/k$, and equilibrium
constants: $X^A_{eq} = f_A/h_A$, $X^R_{eq} = f_R/h_R$, which indicate the ratio between unbinding and binding speed.  There
are four gene states for each gene and the synthesize rates from gene A and B are also symmetric:
$g^A_{ij} = g^B_{ji}$. When gene A is  bound by protein A (self activation) while not bound by protein B (mutual repression), the synthesize
rate of protein for protein A is the largest: $g^A_{01} = F_A + g^A_{11} = F_R+g^A_{00} = F_A+F_R+g^A_{10}$,
where $F_A$ is the activation strength and $F_R$ is the repression strength.
Here, we choose equilibrium constants $X^A_{eq}= X^R_{eq}=45$, symmetric binding/unbinding speed
$\omega_A = \omega_R = \omega$, the repression strength
$F_R=60$ and scale the time to make $k=1$.

\subsection{The potential Landscapes and two mechanisms for cell fate decision of development and differentiation}

Such circuits with above control parameters can generate asymmetric attractors
 representing the differentiated states with almost mutually excluding expression of protein $A$ (i.e. GATA1) and $B$ (i.e. PU.1).
In addition, central symmetric attractor states characterized
by approximately equal levels of $n_A$ and $n_B$ expression can also
be generated, which represent the multipotent state that exhibits the characteristic balanced
or promiscuous expression of the two opposing, fate-determining
concentrations-a hallmark of the indeterminacy of the undecided multipotent stem cell.

We plotted the potential landscape in $n_A$-$n_B$ plane for different activation strength $F_A$ and
binding/unbinding speed $\omega$ in Fig. \ref{fig4}, \ref{fig5}, \ref{fig6},
\ref{fig7}, \ref{fig8}, \ref{fig9}, \ref{fig1}, \ref{fig2}, \ref{fig3} for contour view,
and Fig. \ref{fig4_1}, \ref{fig5_1}, \ref{fig6_1},
\ref{fig7_1}, \ref{fig8_1}, \ref{fig9_1}, \ref{fig1_1}, \ref{fig2_1}, \ref{fig3_1} for 3 dimensional view. In these figures, we found
two kinds of mechanisms for the cell differentiation.

 During the developmental process, the self activation regulation coming from an effective regulation and its change is due to the regulations on these transcription factors mediated by other regulators such as Klf4. When the self activation is strong (large $F_A$), the system is mono-stable with one un-differentiated
central basin, as in Fig. \ref{fig4} (or \ref{fig4_1}). As self activation strength $F_A$ decreases,
the central basin gets weaker and differentiated basins on both sides start to develop,
which results tri-stability as in Fig. \ref{fig7} (or \ref{fig7_1}).
When self activation strength $F_A \rightarrow 0$, the circuit will reduce to a normal symmetric toggle switch.
For toggle switch, $n_A$ and $n_B$ can not be both large in adiabatic limit, because they suppress each other.
Then, the un-differentiated central basin disappeared and two differentiated basins on both sides survives,
which gives bi-stability as in Fig. \ref{fig1} (or \ref{fig1_1}). Therefore, decreasing the self activation
regulatory strength $F_A$ will lead the cell system to differentiate. Changing of the effective self activation regulatory strengths of transcription factors binding to the genes therefore provides a possible differentiation mechanism
which is currently under study \cite{Graf, Huang, Wang2010, Wang2011}.

We would like to point out that there is another possible mechanism of the cell differentiation from the slow binding/unbinding of
protein regulators to gene promoters. We noticed that for a fixed activation
strength $F_A$, cells can develop more stable differentiated states on both sides.
As shown in Fig. \ref{fig4} (or \ref{fig4_1}), \ref{fig5} (or \ref{fig5_1}),
\ref{fig6} (or \ref{fig6_1}) and Fig. \ref{fig7} (or \ref{fig7_1}),
\ref{fig8} (or \ref{fig8_1}), \ref{fig9} (or \ref{fig9_1}), when binding/unbinding rate $\omega$
decreases, the un-differentiated central basin becomes weaker and less stable, while
differentiated basins on both sides become stronger and more stable. We also noticed that
in the non-adiabatic slow binding limit (small binding/unbinding rate $\omega$), multiple meta-stable basins show up. In addition,
in the non-adiabatic slow binding limit, cells have chances to extinct and there are ``extinct basins"
near $(n_A=0, n_B=0)$, as shown in Fig. \ref{fig6} (or \ref{fig6_1}), \ref{fig9} (or \ref{fig9_1}),
\ref{fig3} (or \ref{fig3_1}). These
behaviors are directly due to the non-adiabatic effect: slow binding/unbinding of
protein regulators to promoters. When the binding/unbinding rate $\omega$ is small, the interactions (either repressions or
activations) between gene
states are weak and different gene states statistically co-exist in cells. Each gene state will
give a basin in the concentration and sum of these basins will lead a multiple stable potential
landscape. This results to the development and differentiation with slow binding from the original undifferentiated equally populated single basin of attraction with fast binding. Slow binding provides another possible mechanism for differentiation and development.

\subsection{Kinetic and optimal speed for development and differentiation}

To quantitatively characterize the dynamics of the differentiation and the reverse process as reprogramming, we study the speed of differentiation and reprogramming in terms of mean first passage time (MFPT) , as shown in Fig. \ref{fig11}, \ref{fig12} and \ref{fig13}.
In an attractor landscape, the lifetime
of an attractor reflects its stability, which can be measured by MFPT. MFPT is the average transition
time induced by intrinsic statistical fluctuations of molecule numbers between attractors on a landscape, since the
traversing time represents how easy to switch from one place
to another. When the binding/unbinding rate $\omega$ is relatively large, the un-differentiated central
basin becomes more stable, as in Fig. \ref{fig4} (or \ref{fig4_1}),
\ref{fig7} (or \ref{fig7_1}), \ref{fig1} (or \ref{fig1_1}),
and cells have more chances to stay in the un-differentiated state.
Therefore, the differentiation process will be more difficult and MFPT is longer for faster binding.
For the differentiation process, it is noticed that, as the binding/unbinding
rate $\omega$ increases, the MFPT decreases first, and then increases. In the non-adiabatic limit (small binding/unbinding rate $\omega$),
the differentiation limiting step is the
binding/unbinding events. Therefore, increasing binding/unbinding speed $\omega$ will accelerate
the kinetics from the un-differentiated central basin to differentiated side basins. So for the differentiation process, caused from faster binding to slower binding of regulatory proteins to the genes, we notice that the speed for differentiation is slow when state is dominated by undifferentiated state for faster binding, and is also slower for slower binding which is due to the occasional binding being the rate limiting step for differentiation. There is an optimal speed for differentiation. As binding becomes faster from low speed end (non-adiabatic limit), the speed of differentiation is controlled by the binding speed and therefore increases. As the binding becomes even faster, the differentiation is dominated by the escape from the undifferentiated basin of attraction and therefore is significantly slowed down. This creates an optimal speed for differentiation and development.

The reverse process of cell differentiation is the reprogramming of differentiated
cells back to a multi- or pluripotent state. In Fig. \ref{fig11}, \ref{fig12} and \ref{fig13},
the MFPT for the reprogramming for different self activation strength $F_A$ and binding/unbinding speed $\omega$ is plotted.
We observed that, for a typical differentiated system,
as in Fig. \ref{fig6} (or \ref{fig6_1}) and Fig. \ref{fig1} (or \ref{fig1_1}), the reprogramming chance is very low and
requires a very long time. For self activation strength $F_A =20$ (Fig. \ref{fig11}) and $F_A =13$ (\ref{fig12}),
the MFPT for the reprogramming decrease as the increasing of thbinding/unbinding speed $\omega$, because
the stability of un-differentiated symmetric central state increases with the binding/unbinding speed $\omega$ as
we can see in potential landscapes, Fig. \ref{fig4} (or \ref{fig4_1}), \ref{fig5} (or \ref{fig5_1}),
\ref{fig6} (or \ref{fig6_1}), \ref{fig7} (or \ref{fig7_1}), \ref{fig8} (or \ref{fig8_1}) and
\ref{fig9} (or \ref{fig9_1}). While in Fig. \ref{fig13}, since there is no
self-activation and no stable symmetric central basin in the landscape,
the reprogramming is difficult and the MFPT is very long for different  the binding/unbinding speed $\omega$.

\subsection{Biological dynamic pathways of differentiation and reprogramming}

Both the differentiation and reprogramming can be caused by the change of
gene regulations during the developmental process. Here we consider the evolution of the binding/unbinding rate from fast to slow
$\omega (t) = 1000 e^{-\kappa t} + 0.001$ from $1000 \rightarrow 0.001$ for the differentiation and
the evolution of the binding/unbinding rate
$\omega(t) = 1000 [1-e^{-\kappa t} ] + 0.001$ from $0.001 \rightarrow 1000$ for the reprogramming from slow to fast.
The transition paths from Gillespie simulation are plotted in Fig. \ref{path}, accompanied with
the potential landscapes for the binding/unbinding speed $\omega=0.001, 1, 1000$. It is interesting to observe that
the biological dynamic paths  are irreversible, i.e. the differentiation path and reprogramming
path are totally different. In the differentiation process, the system stay on the multipotent undifferentiated state
for a while until binding becomes slower. As the binding becomes slower, the undifferentiated state becomes less stable.
Furthermore, the gene state can be switched through binding/unbinding event of
regulatory proteins to the promoters and the system will then be evolved from the undifferentiated basin to
the differentiated basin of attraction. In the reprogramming process,
the system will be gradually attracted into the undifferentiated basin as the increasing of the
binding/unbinding rate $\omega$. The paths of differentiation do not
follow the gradient steepest descent of the potential landscape. They do
not follow the paths of the reprogramming (the reverse differentiation process). This  irreversibility reflects the
underlying non-equilibrium nature of the differentiation and developmental network systems \cite{Wang2011}.
It can give us the fundamental understanding of the biological origin of time arrow in cell development.

\section{Conclusion}

We developed a theoretical framework to quantify the potential
landscape and biological paths for cell development and differentiation.
We found a new mechanism for differentiation. The differentiated state can emerge from
the slow promoter binding/unbinding processes. We found under slow promoter binding, there can be
many meta-stable differentiated states. This has been observed experimentally \cite{Arias}.
Our theory gives a possible explanation for the origins of those meta-stable states in the experiments.

We show that the developmental process can be quantitatively
described and uncovered by the biological paths on the
potential landscape and the dynamics of the developmental
process is controlled by a combination of the intrinsic fluctuations of
protein concentrations and gene state fluctuations through promoter binding.
We also show that
the biological paths of the reverse differentiation process
or reprogramming are irreversible and different from the
ones of the differentiation process.

We explored the kinetic speed for differentiation. We found that the cell differentiation and reprogramming
dynamics strongly depends on the binding/unbinding rate of the regulatory proteins
to the gene promoters. We found an optimal speed for differentiation and development with certain binding/unbinding rates of
regulatory proteins to the gene promoters.
An interesting question we may ask is that is the differentiation and development at optimal speed?
More experimental and bioinformatics studies might be able to pin down the answer.
Furthermore, the irreversibility in cell development gives biological examples, which can be
easily observed in experiments,
for the understanding of the origin of time arrow in general non-equilibrium systems.

\newpage

\setcounter{equation}{0}
\renewcommand{\theequation}{{S}\arabic{equation}}
\begin{center}
\textbf{Supplementary Informations: Master Equations}
\end{center}

\begin{figure}[ht]
\begin{center}
\hspace{0mm} \includegraphics[scale=0.5]{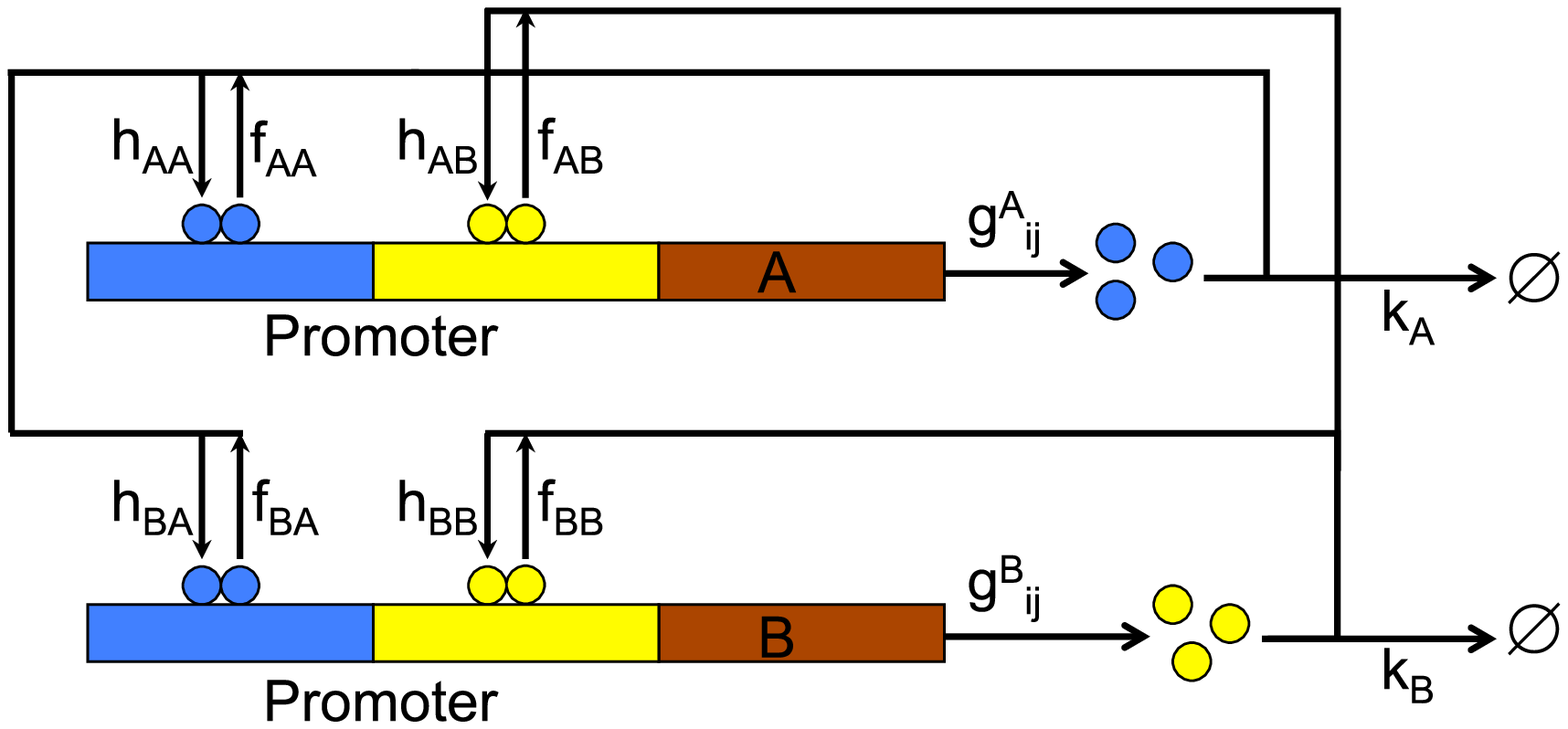}
\end{center}
\caption{\label{circuit} Network diagram of canonical gene regulatory circuit of two mutually opposing proteins that positively self-regulate themselves.
Two types of genes, $A$ and $B$ are translated into proteins $A$
and $B$ respectively. The proteins $A$($B$) can bind to the
promoter of the gene $A$($B$) to activate the synthesis rate of
$A$($B$), which makes a self-activation feedback loop.
The proteins $A$($B$) can bind to the gene $B$($A$) to
repress the synthesis rate of $B$($A$), which makes a
mutual repression loop. Both protein $A$ and protein $B$ bind on promoters
as a dimer with the binding rate $h_{\alpha A} = \frac{1}{2} h_{\alpha A} n_A (n_A -1)$,
$h_{\alpha B}=\frac{1}{2} h_{\alpha B} n_B (n_B-1)$ respectively, and
the unbinding rate $f_{\alpha A}$, $f_{\alpha B}$ respectively, with $\alpha = (A, B)$.}
\end{figure}

\begin{figure}
\begin{center}
\subfigure[$F_A=20, \omega=1000$]{
\hspace{0mm}\includegraphics[width=0.3\columnwidth]{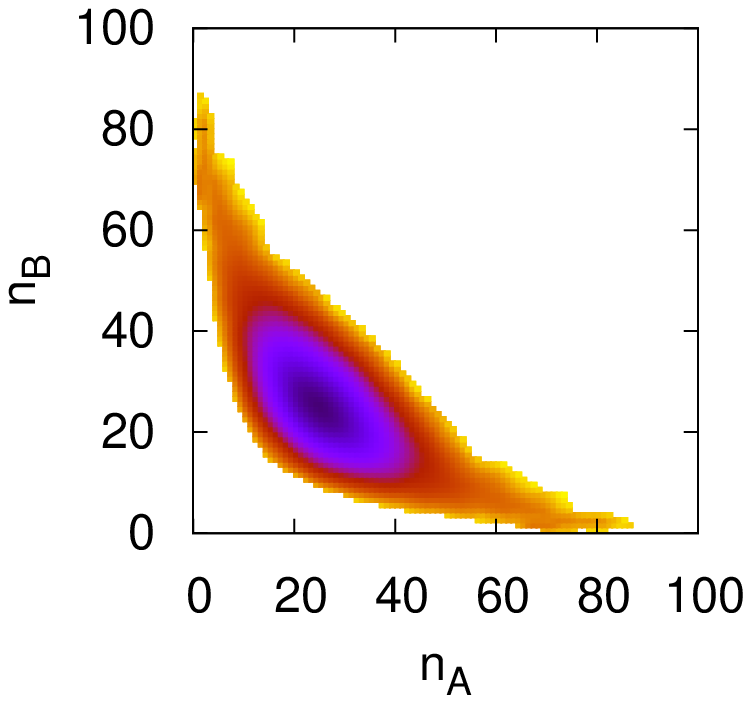} \label{fig4} }
\subfigure[$F_A=20, \omega=1$]{
\hspace{0mm}\includegraphics[width=0.3\columnwidth]{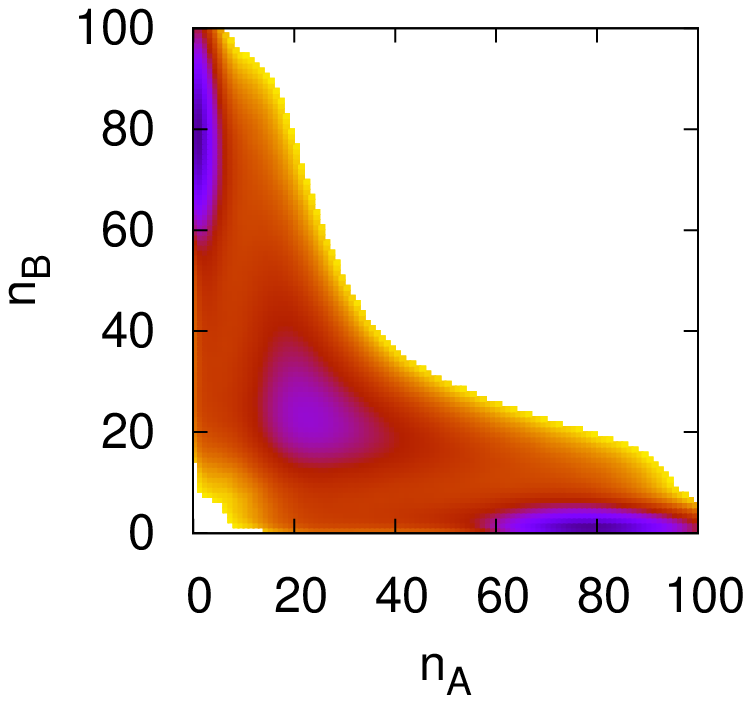} \label{fig5} }
\subfigure[$F_A=20, \omega=0.001$]{
\hspace{0mm}\includegraphics[width=0.3\columnwidth]{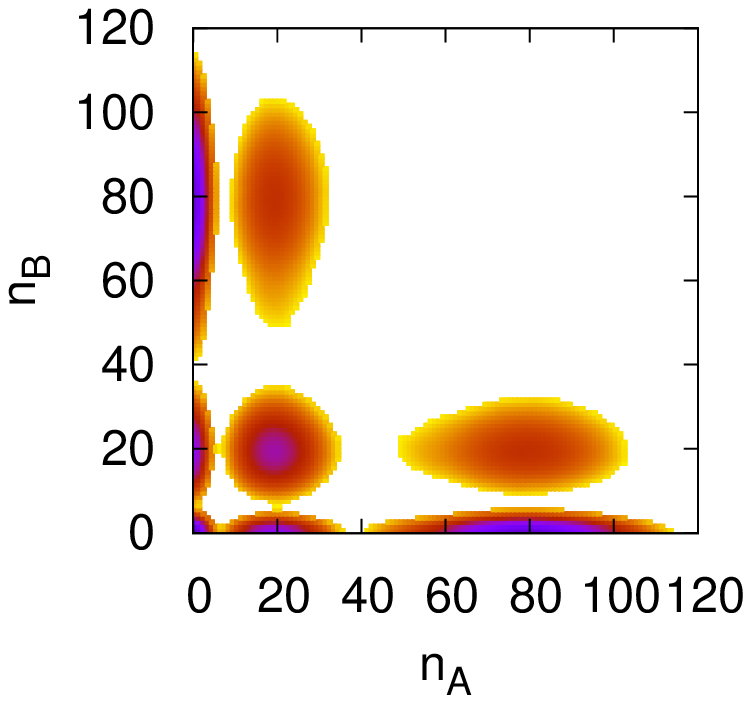} \label{fig6} }
\subfigure[$F_A=13, \omega=1000$]{
\hspace{0mm}\includegraphics[width=0.3\columnwidth]{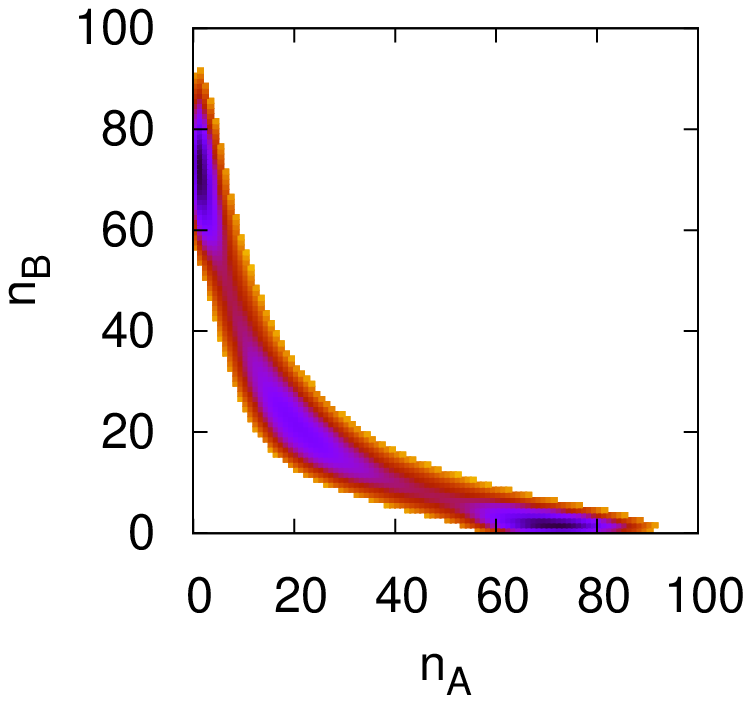} \label{fig7} }
\subfigure[$F_A=13, \omega=1$]{
\hspace{0mm}\includegraphics[width=0.3\columnwidth]{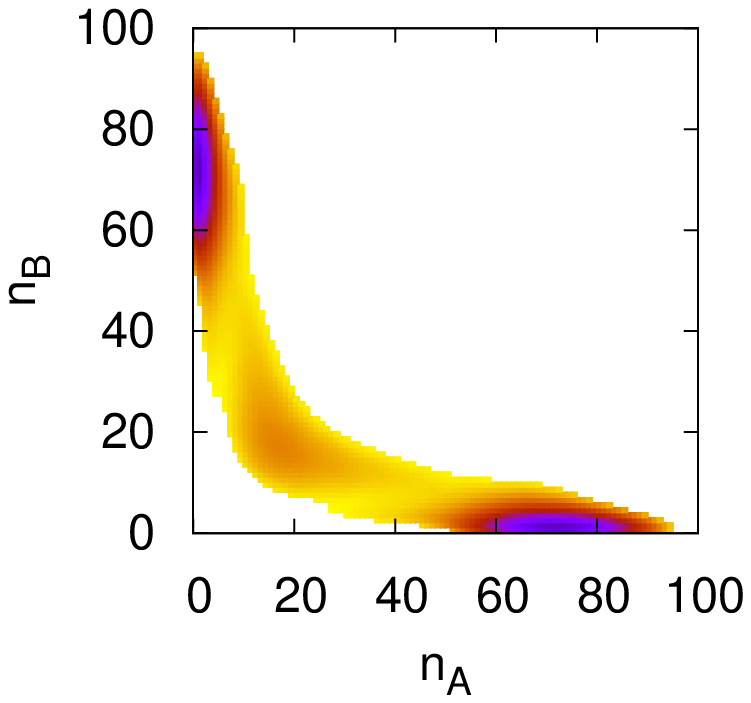} \label{fig8} }
\subfigure[$F_A=13, \omega=0.001$]{
\hspace{0mm}\includegraphics[width=0.3\columnwidth]{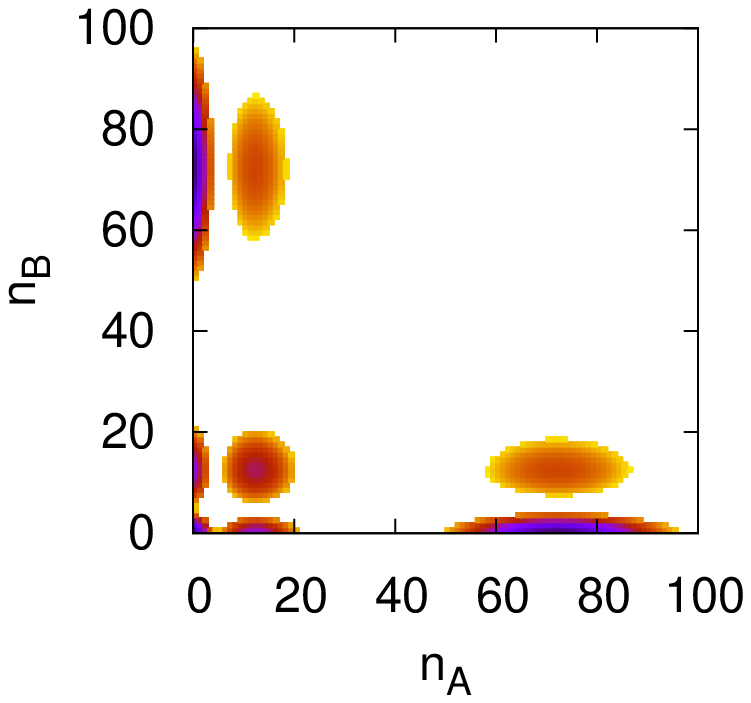} \label{fig9} }
\subfigure[$F_A=0, \omega=1000$]{
\hspace{0mm}\includegraphics[width=0.3\columnwidth]{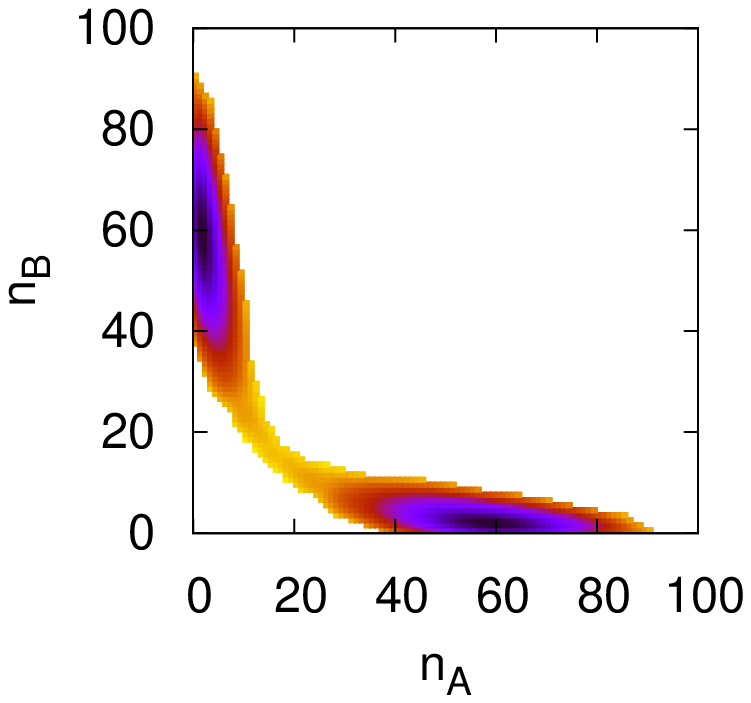} \label{fig1} }
\subfigure[$F_A=0, \omega=1$]{
\hspace{0mm}\includegraphics[width=0.3\columnwidth]{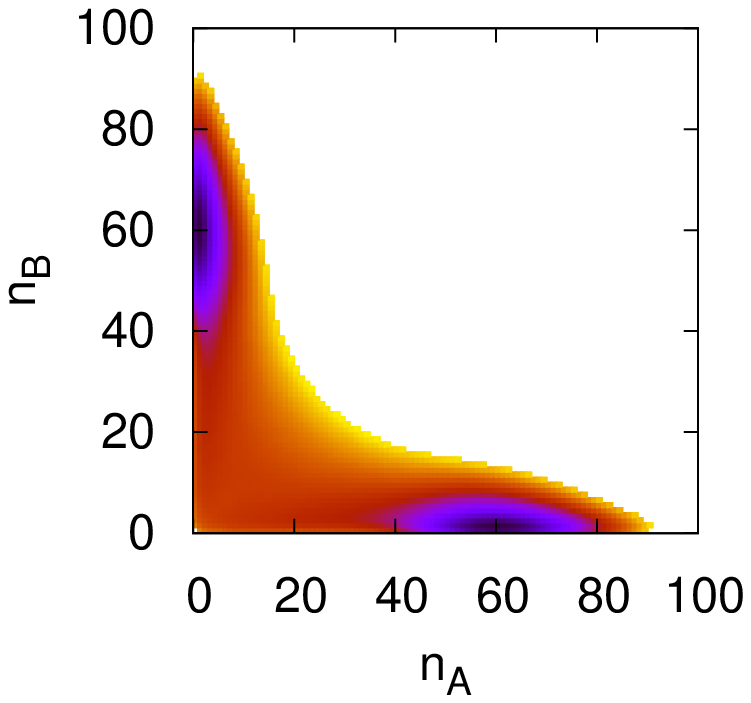} \label{fig2} }
\subfigure[$F_A=0, \omega=0.001$]{
\hspace{0mm}\includegraphics[width=0.3\columnwidth]{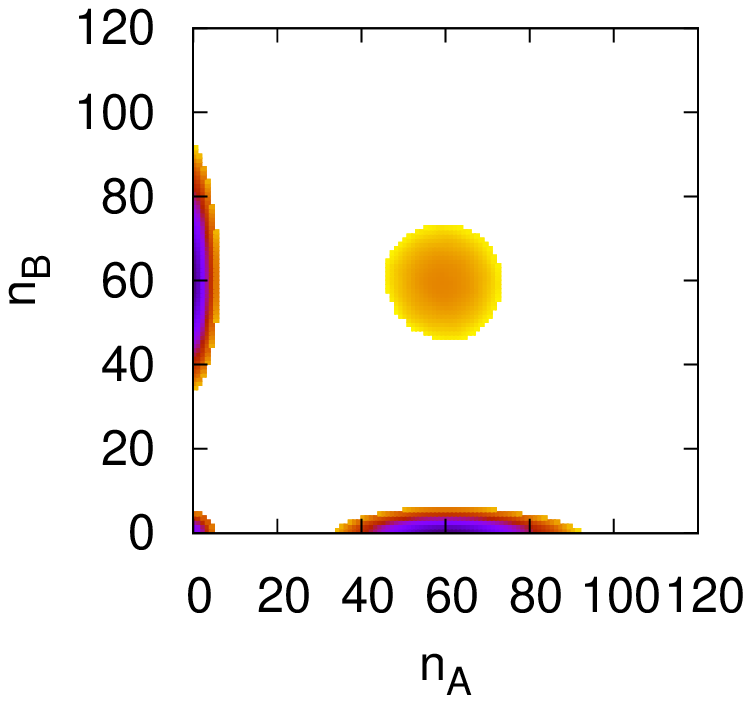} \label{fig3} }
\end{center}
\caption[Optional caption for list of
figures]{The potential landscape (contour view) in $n_A$-$n_B$ plane for different self activation strength $F_A$
and binding/unbinding speed $\omega$.
Differentiation happens with the decrease of the binding/unbinding speed $\omega$ (from left to right) or the decrease of
the activation strength $F_A$ (from top to bottom).}
\end{figure}

\begin{figure}
\begin{center}
\subfigure[$F_A=20, \omega=1000$]{
\hspace{0mm}\includegraphics[width=0.26\columnwidth]{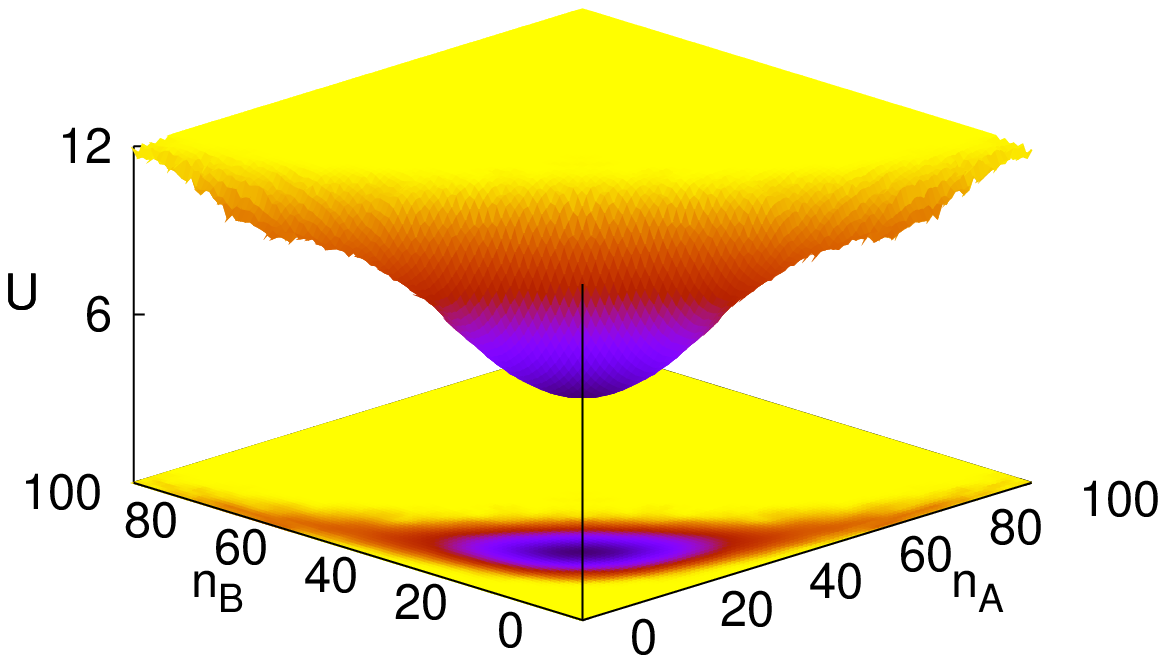} \label{fig4_1} }
\subfigure[$F_A=20, \omega=1$]{
\hspace{0mm}\includegraphics[width=0.26\columnwidth]{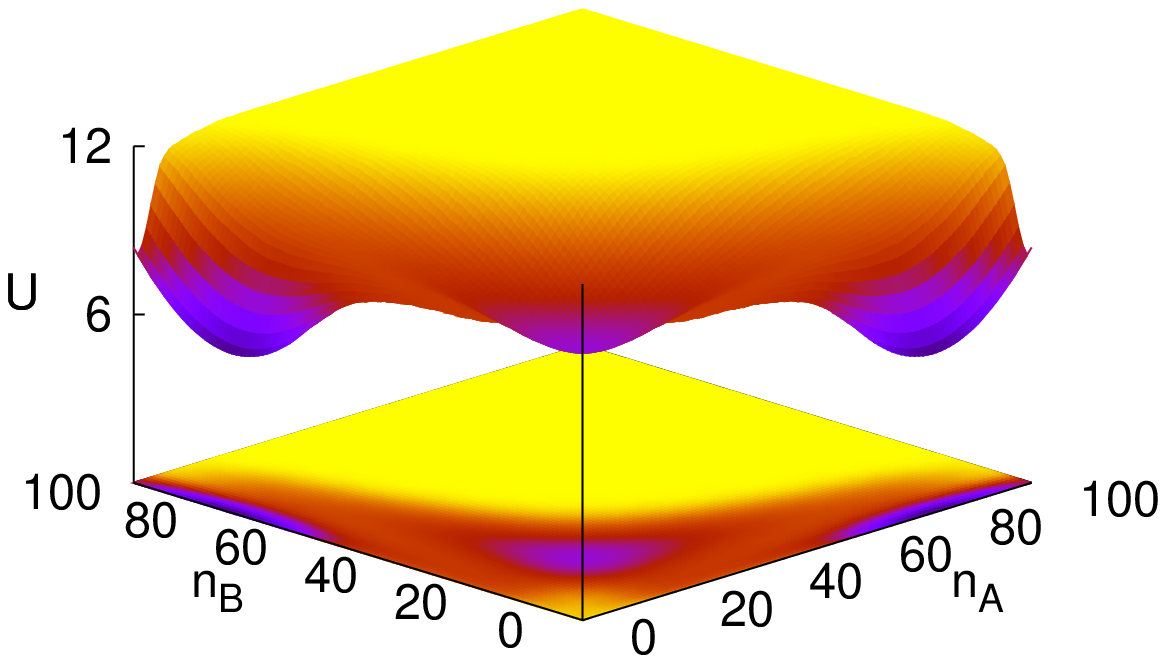} \label{fig5_1} }
\subfigure[$F_A=20, \omega=0.001$]{
\hspace{0mm}\includegraphics[width=0.26\columnwidth]{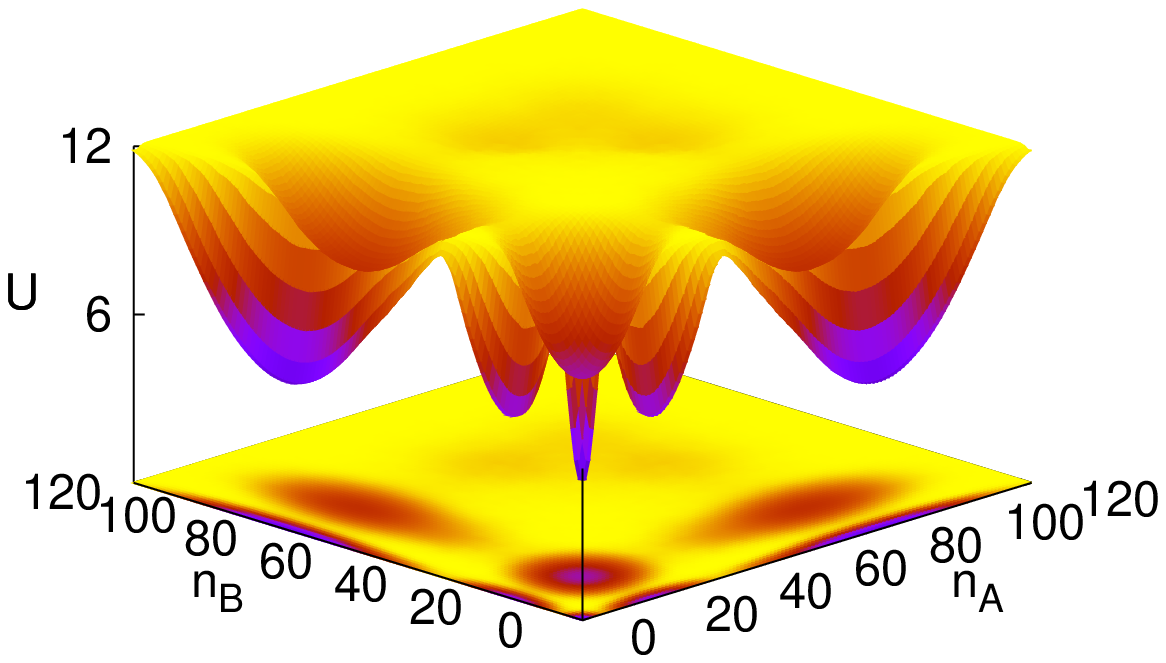} \label{fig6_1} }
\subfigure[$F_A=13, \omega=1000$]{
\hspace{0mm}\includegraphics[width=0.26\columnwidth]{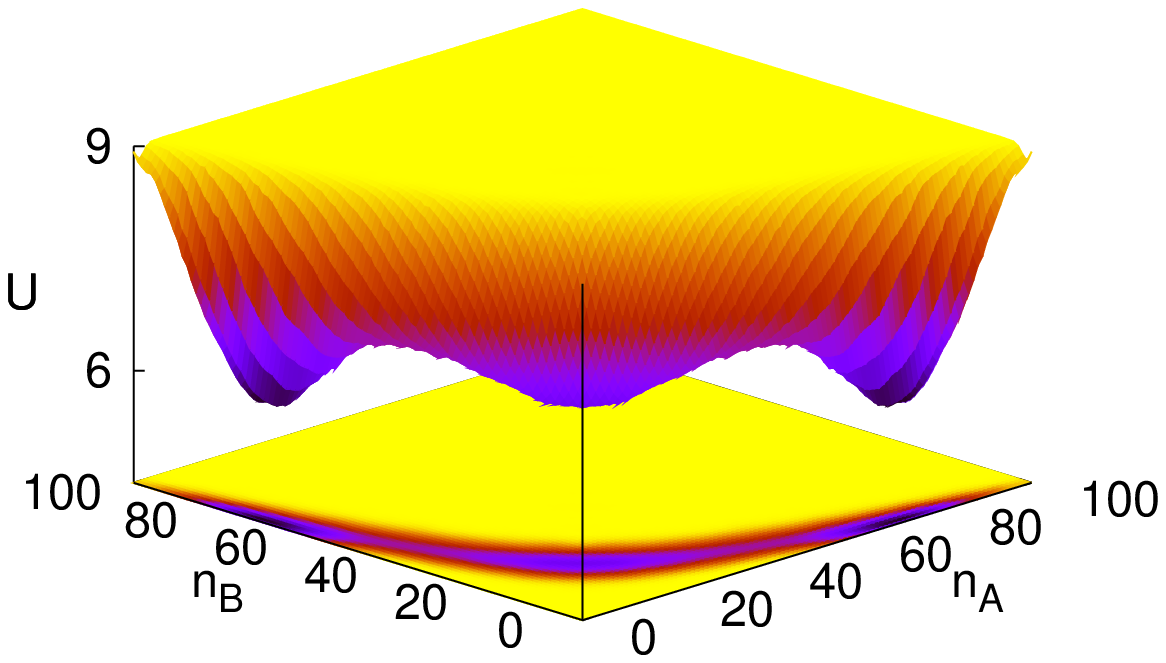} \label{fig7_1} }
\subfigure[$F_A=13, \omega=1$]{
\hspace{0mm}\includegraphics[width=0.26\columnwidth]{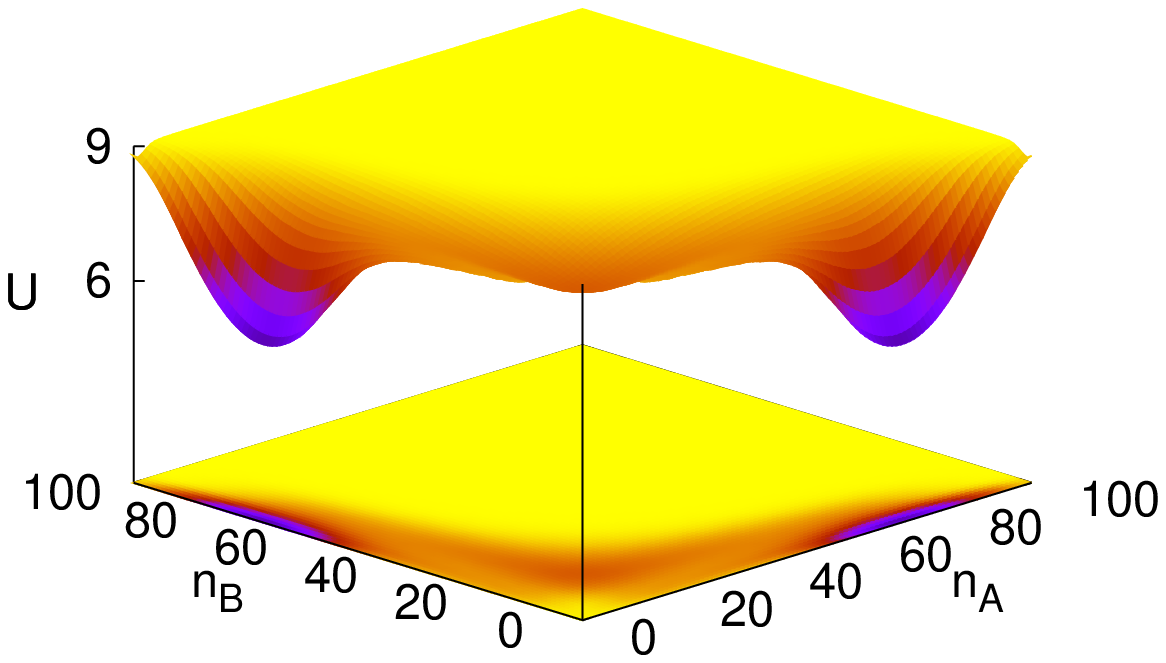} \label{fig8_1} }
\subfigure[$F_A=13, \omega=0.001$]{
\hspace{0mm}\includegraphics[width=0.26\columnwidth]{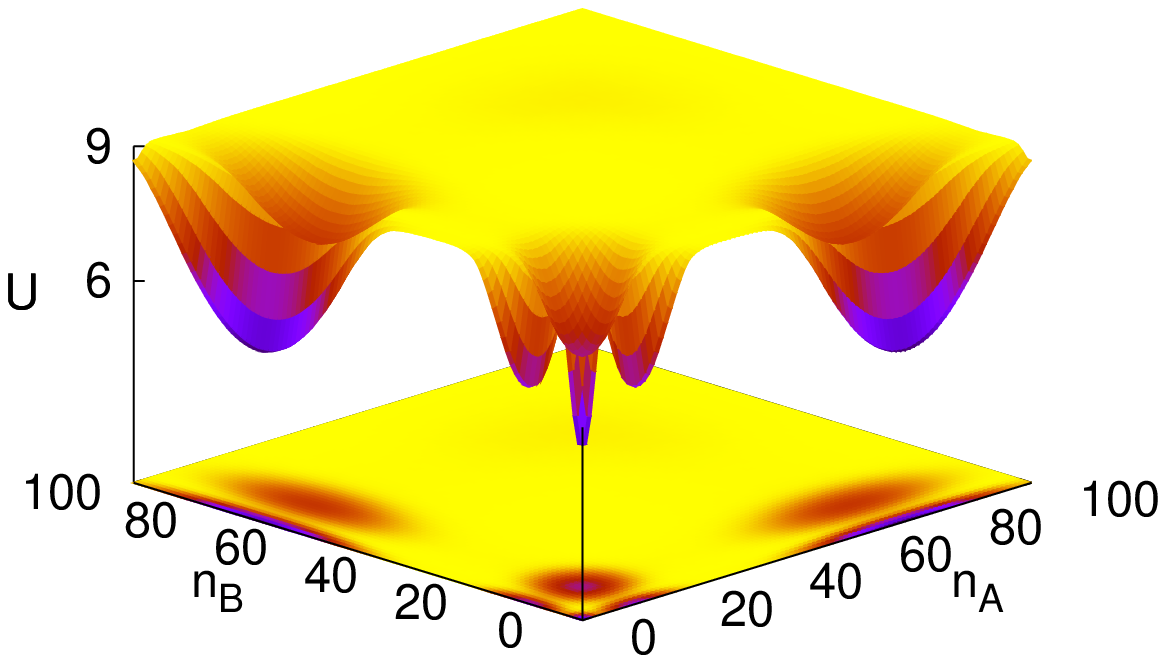} \label{fig9_1} }
\subfigure[$F_A=0, \omega=1000$]{
\hspace{0mm}\includegraphics[width=0.26\columnwidth]{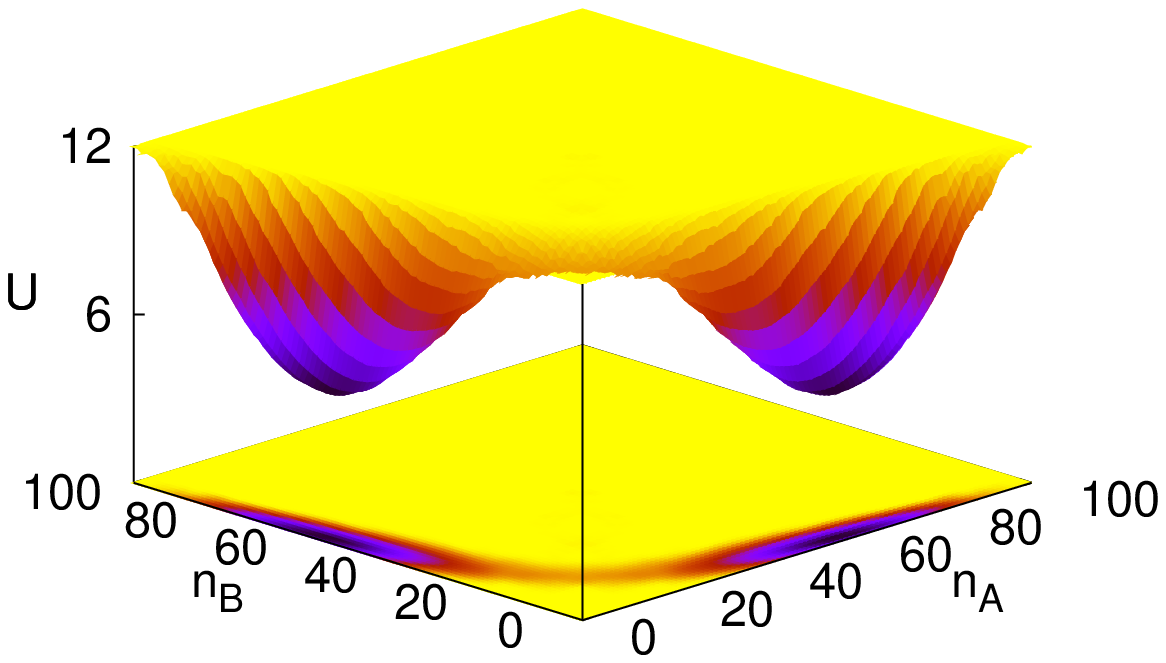} \label{fig1_1} }
\subfigure[$F_A=0, \omega=1$]{
\hspace{0mm}\includegraphics[width=0.26\columnwidth]{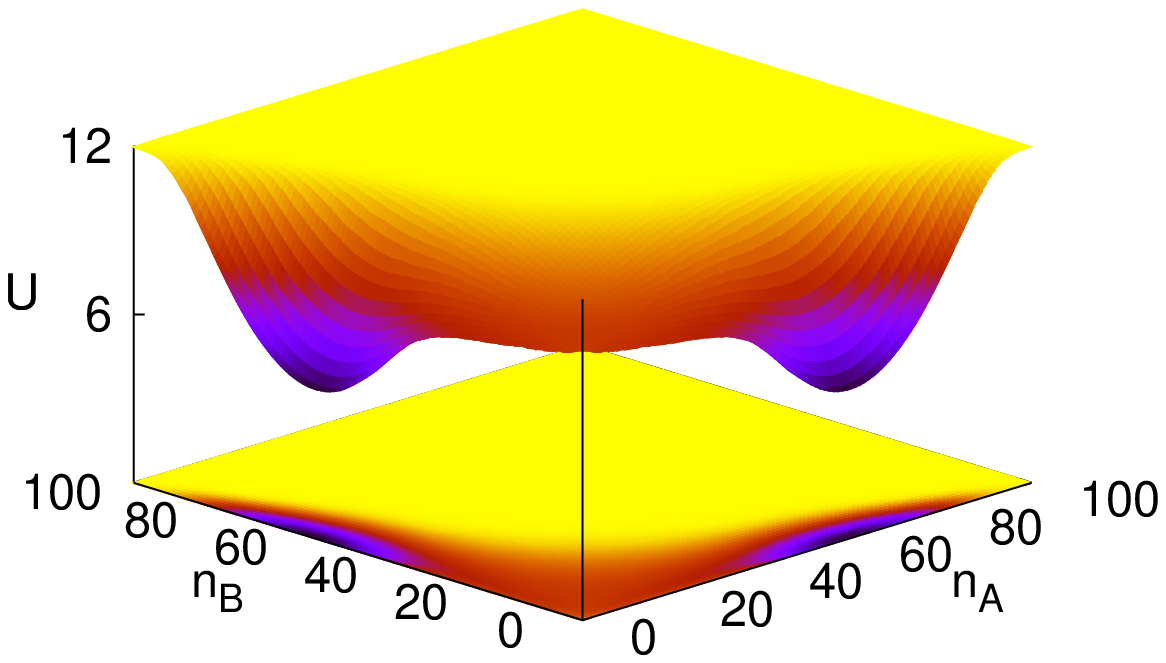} \label{fig2_1} }
\subfigure[$F_A=0, \omega=0.001$]{
\hspace{0mm}\includegraphics[width=0.26\columnwidth]{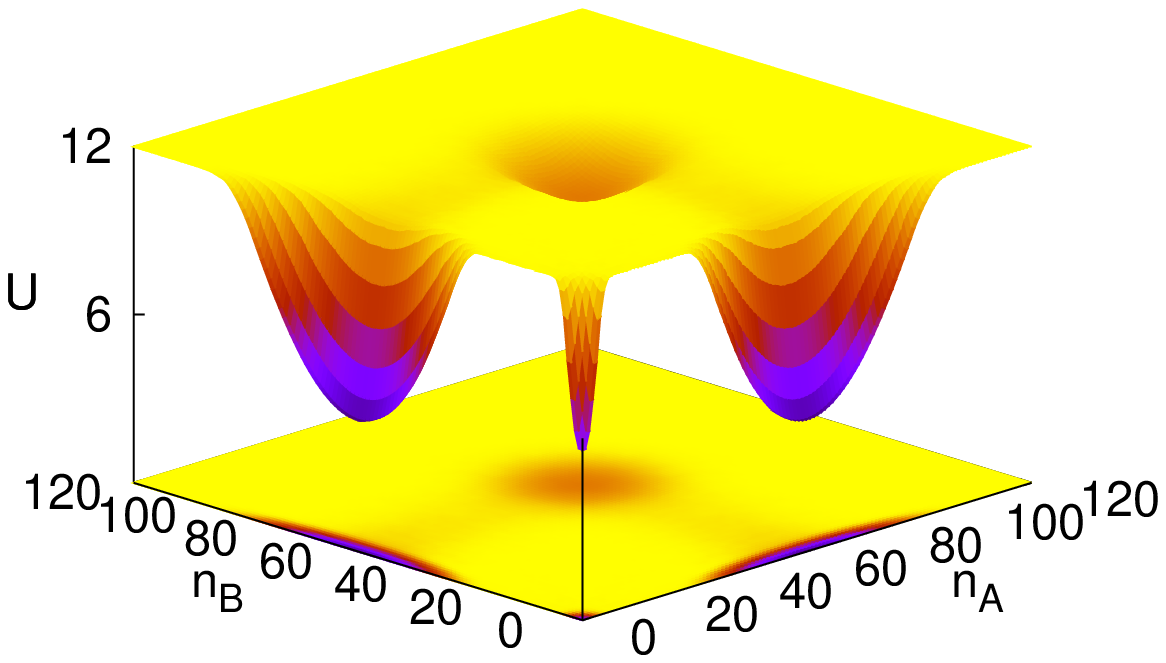} \label{fig3_1} }
\end{center}
\caption[Optional caption for list of
figures]{The potential landscape (3 dimensional view) in $n_A$-$n_B$ plane
for different self activation strength $F_A$ and binding/unbinding speed $\omega$. Differentiation happens
with the decrease of the binding/unbinding speed $\omega$ (from left to right) or the decrease of
the activation strength $F_A$ (from top to bottom).}
\end{figure}

\begin{figure}
\begin{center}
\subfigure[$F_A = 20$]{
\hspace{0mm}\includegraphics[width=0.55\columnwidth]{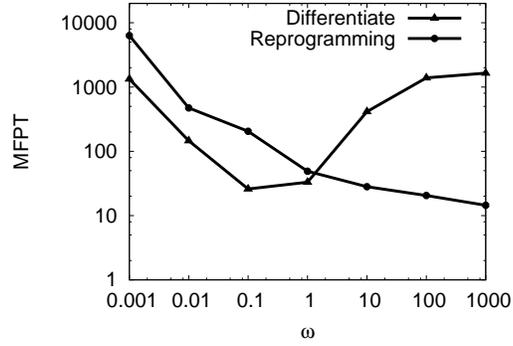} \label{fig11} }
\subfigure[$F_A = 13$]{
\hspace{0mm}\includegraphics[width=0.55\columnwidth]{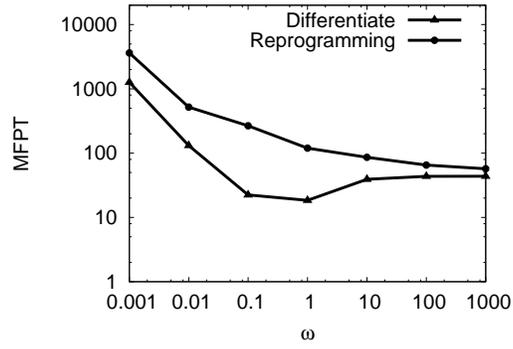} \label{fig12} }
\subfigure[$F_A = 0$]{
\hspace{0mm}\includegraphics[width=0.55\columnwidth]{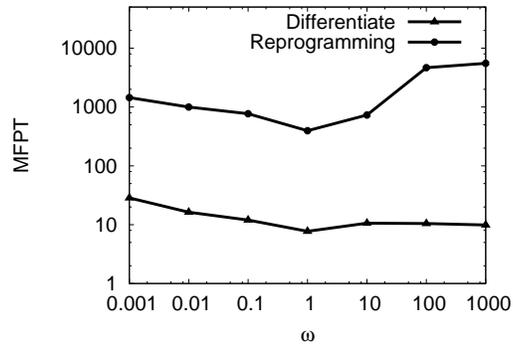} \label{fig13} }
\end{center}
\caption[Optional caption for list of
figures]{The MFPT of the differentiation and reprogramming for different self activation strength $F_A$
and binding/unbinding speed $\omega$. }
\end{figure}

\begin{figure}[ht]
\begin{center}
\hspace{10mm} \includegraphics[scale=0.9]{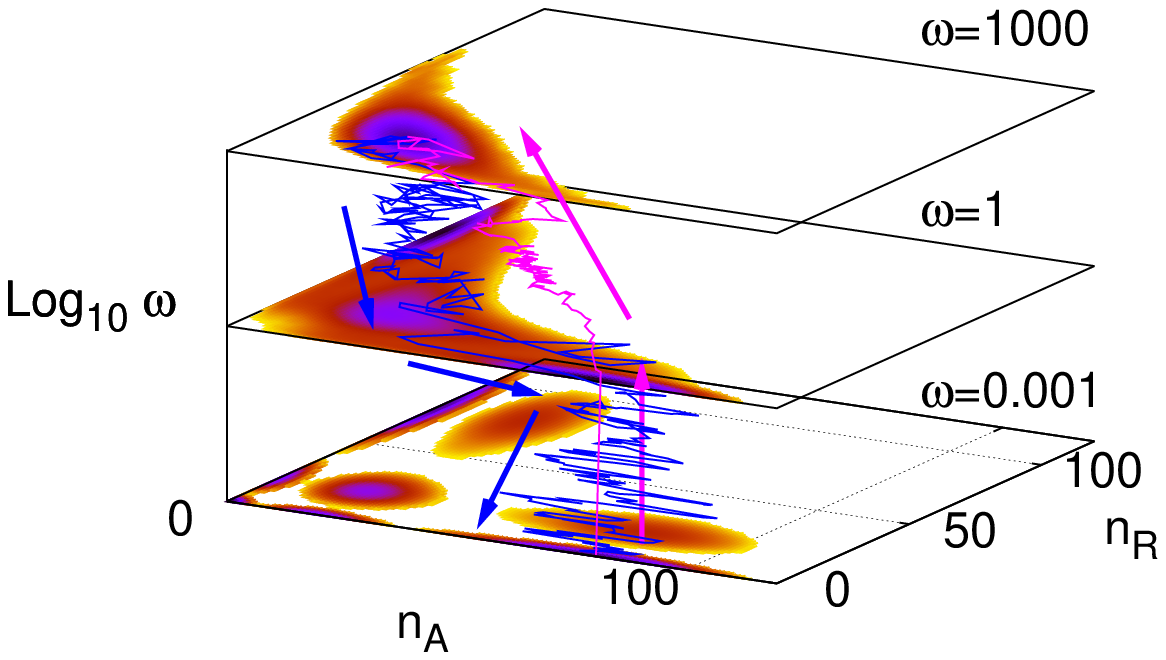}
\end{center}
\caption{\label{path} Transition paths for differentiation (blue) and reprogramming (purple) with $\kappa=0.1$,
with self activation strength $F_A=20$. }
\end{figure}

16 Master equations for the canonical gene regulatory circuit of two mutually opposing proteins that positively self-regulate themselves, as in Fig. \ref{circuit},
are given as following:

\begin{eqnarray}\label{MS1}
\frac{d P_{1111} (n_{A}, n_{B})}{dt} = \nonumber \\
-\frac{h_{AA}}{2} [n_{A} (n_{A}-1)] P_{1111} (n_{A}, n_{B}) + f_{AA} P_{0111} (n_{A}-2, n_{B}) \nonumber \\
-\frac{h_{AB}}{2} [n_{B} (n_{B}-1)] P_{1111} (n_{A}, n_{B}) + f_{AB} P_{1011} (n_{A}, n_{B}-2) \nonumber \\
-\frac{h_{BA}}{2} [n_{A} (n_{A}-1)] P_{1111} (n_{A}, n_{B}) + f_{BA} P_{1101} (n_{A}-2, n_{B}) \nonumber \\
-\frac{h_{BB}}{2} [n_{B} (n_{B}-1)] P_{1111} (n_{A}, n_{B}) + f_{BB} P_{1110} (n_{A}, n_{B}-2) \nonumber \\
+k_{A} [(n_{A} +1) P_{1111} (n_{A}+1, n_{B}) - n_{A} P_{1111} (n_{A}, n_{B})] \nonumber \\
+k_{B} [(n_{B} +1) P_{1111} (n_{A}, n_{B}+1) - n_{B} P_{1111} (n_{A}, n_{B})] \nonumber \\
+g^A_{11} [P_{1111} (n_{A}-1, n_{B}) - P_{1111} (n_{A}, n_{B})] \nonumber \\
+g^B_{11} [P_{1111} (n_{A}, n_{B}-1) - P_{1111} (n_{A}, n_{B})]
\end{eqnarray}

\begin{eqnarray}
\frac{d P_{1011} (n_{A}, n_{B})}{dt} = \nonumber \\
- \frac{h_{AA}}{2} [n_{A} (n_{A}-1)] P_{1011} (n_{A}, n_{B}) + f_{AA} P_{0011} (n_{A}-2, n_{B}) \nonumber \\
+ \frac{h_{AB}}{2} [(n_{B}+2) (n_{B}+1)] P_{1111} (n_{A}, n_{B}+2) - f_{AB} P_{1011} (n_{A}, n_{B}) \nonumber \\
-\frac{h_{BA}}{2} [n_{A} (n_{A}-1)] P_{1011} (n_{A}, n_{B}) + f_{BA} P_{1001} (n_{A}-2, n_{B}) \nonumber \\
-\frac{h_{BB}}{2} [n_{B} (n_{B}-1)] P_{1011} (n_{A}, n_{B}) + f_{BB} P_{1010} (n_{A}, n_{B}-2) \nonumber \\
+k_{A} [(n_{A} +1) P_{1011} (n_{A}+1, n_{B}) - n_{A} P_{1011} (n_{A}, n_{B})] \nonumber \\
+k_{B} [(n_{B} +1) P_{1011} (n_{A}, n_{B}+1) - n_{B} P_{1011} (n_{A}, n_{B})] \nonumber \\
+g^A_{10} [P_{1011} (n_{A}-1, n_{B}) - P_{1011} (n_{A}, n_{B})] \nonumber \\
+g^B_{11} [P_{1011} (n_{A}, n_{B}-1) - P_{1011} (n_{A}, n_{B})]
\end{eqnarray}

\begin{eqnarray}
\frac{d P_{0111} (n_{A}, n_{B})}{dt} = \nonumber \\
+\frac{h_{AA}}{2} [(n_{A}+2) (n_{A}+1)] P_{1111} (n_{A}+2, n_{B}) - f_{AA} P_{0111} (n_{A}, n_{B}) \nonumber \\
-\frac{h_{AB}}{2} [n_{B} (n_{B}-1)] P_{0111} (n_{A}, n_{B}) + f_{AB} P_{0011} (n_{A}, n_{B}-2) \nonumber \\
-\frac{h_{BA}}{2} [n_{A} (n_{A}-1)] P_{0111} (n_{A}, n_{B}) + f_{BA} P_{0101} (n_{A}-2, n_{B}) \nonumber \\
-\frac{h_{BB}}{2} [n_{B} (n_{B}-1)] P_{0111} (n_{A}, n_{B}) + f_{BB} P_{0110} (n_{A}, n_{B}-2) \nonumber \\
+k_{A} [(n_{A} +1) P_{0111} (n_{A}+1, n_{B}) - n_{A} P_{0111} (n_{A}, n_{B})] \nonumber \\
+k_{B} [(n_{B} +1) P_{0111} (n_{A}, n_{B}+1) - n_{B} P_{0111} (n_{A}, n_{B})] \nonumber \\
+g^A_{01} [P_{0111} (n_{A}-1, n_{B}) - P_{0111} (n_{A}, n_{B})] \nonumber \\
+g^B_{11} [P_{0111} (n_{A}, n_{B}-1) - P_{0111} (n_{A}, n_{B})]
\end{eqnarray}

\begin{eqnarray}
\frac{d P_{0011} (n_{A}, n_{B})}{dt} = \nonumber \\
+ \frac{h_{AA}}{2} [(n_{A}+2) (n_{A}+1)] P_{1011} (n_{A}+2, n_{B}) - f_{AA} P_{0011} (n_{A}, n_{B}) \nonumber \\
+ \frac{h_{AB}}{2} [(n_{B}+2) (n_{B}+1)] P_{0111} (n_{A}, n_{B}+2) - f_{AB} P_{0011} (n_{A}, n_{B}) \nonumber \\
-\frac{h_{BA}}{2} [n_{A} (n_{A}-1)] P_{0011} (n_{A}, n_{B}) + f_{BA} P_{0001} (n_{A}-2, n_{B}) \nonumber \\
-\frac{h_{BB}}{2} [n_{B} (n_{B}-1)] P_{0011} (n_{A}, n_{B}) + f_{BB} P_{0010} (n_{A}, n_{B}-2) \nonumber \\
+k_{A} [(n_{A} +1) P_{0011} (n_{A}+1, n_{B}) - n_{A} P_{0011} (n_{A}, n_{B})] \nonumber \\
+k_{B} [(n_{B} +1) P_{0011} (n_{A}, n_{B}+1) - n_{B} P_{0011} (n_{A}, n_{B})] \nonumber \\
+g^A_{00} [P_{0011} (n_{A}-1, n_{B}) - P_{0011} (n_{A}, n_{B})] \nonumber \\
+g^B_{11} [P_{0011} (n_{A}, n_{B}-1) - P_{0011} (n_{A}, n_{B})]
\end{eqnarray}

\begin{eqnarray}
\frac{d P_{1110} (n_{A}, n_{B})}{dt} = \nonumber \\
-\frac{h_{AA}}{2} [n_{A} (n_{A}-1)] P_{1110} (n_{A}, n_{B}) + f_{AA} P_{0110} (n_{A}-2, n_{B}) \nonumber \\
-\frac{h_{AB}}{2} [n_{B} (n_{B}-1)] P_{1110} (n_{A}, n_{B}) + f_{AB} P_{1010} (n_{A}, n_{B}-2) \nonumber \\
-\frac{h_{BA}}{2} [n_{A} (n_{A}-1)] P_{1110} (n_{A}, n_{B}) + f_{BA} P_{1100} (n_{A}-2, n_{B}) \nonumber \\
+\frac{h_{BB}}{2} [(n_{B}+2) (n_{B}+1)] P_{1111} (n_{A}, n_{B}+2) - f_{BB} P_{1110} (n_{A}, n_{B}) \nonumber \\
+k_{A} [(n_{A} +1) P_{1110} (n_{A}+1, n_{B}) - n_{A} P_{1110} (n_{A}, n_{B})] \nonumber \\
+k_{B} [(n_{B} +1) P_{1110} (n_{A}, n_{B}+1) - n_{B} P_{1110} (n_{A}, n_{B})] \nonumber \\
+g^A_{11} [P_{1110} (n_{A}-1, n_{B}) - P_{1110} (n_{A}, n_{B})] \nonumber \\
+g^B_{10} [P_{1110} (n_{A}, n_{B}-1) - P_{1110} (n_{A}, n_{B})]
\end{eqnarray}

\begin{eqnarray}
\frac{d P_{1010} (n_{A}, n_{B})}{dt} = \nonumber \\
- \frac{h_{AA}}{2} [n_{A} (n_{A}-1)] P_{1010} (n_{A}, n_{B}) + f_{AA} P_{0010} (n_{A}-2, n_{B}) \nonumber \\
+ \frac{h_{AB}}{2} [(n_{B}+2) (n_{B}+1)] P_{1110} (n_{A}, n_{B}+2) - f_{AB} P_{1010} (n_{A}, n_{B}) \nonumber \\
-\frac{h_{BA}}{2} [n_{A} (n_{A}-1)] P_{1010} (n_{A}, n_{B}) + f_{BA} P_{1000} (n_{A}-2, n_{B}) \nonumber \\
+\frac{h_{BB}}{2} [(n_{B}+2) (n_{B}+1)] P_{1011} (n_{A}, n_{B}+2) - f_{BB} P_{1010} (n_{A}, n_{B}) \nonumber \\
+k_{A} [(n_{A} +1) P_{1010} (n_{A}+1, n_{B}) - n_{A} P_{1010} (n_{A}, n_{B})] \nonumber \\
+k_{B} [(n_{B} +1) P_{1010} (n_{A}, n_{B}+1) - n_{B} P_{1010} (n_{A}, n_{B})] \nonumber \\
+g^A_{10} [P_{1010} (n_{A}-1, n_{B}) - P_{1010} (n_{A}, n_{B})] \nonumber \\
+g^B_{10} [P_{1010} (n_{A}, n_{B}-1) - P_{1010} (n_{A}, n_{B})]
\end{eqnarray}

\begin{eqnarray}
\frac{d P_{0110} (n_{A}, n_{B})}{dt} = \nonumber \\
+\frac{h_{AA}}{2} [(n_{A}+2) (n_{A}+1)] P_{1110} (n_{A}+2, n_{B}) - f_{AA} P_{0110} (n_{A}, n_{B}) \nonumber \\
-\frac{h_{AB}}{2} [n_{B} (n_{B}-1)] P_{0110} (n_{A}, n_{B}) + f_{AB} P_{0010} (n_{A}, n_{B}-2) \nonumber \\
-\frac{h_{BA}}{2} [n_{A} (n_{A}-1)] P_{0110} (n_{A}, n_{B}) + f_{BA} P_{0100} (n_{A}-2, n_{B}) \nonumber \\
+\frac{h_{BB}}{2} [(n_{B}+2) (n_{B}+1)] P_{0111} (n_{A}, n_{B}+2) - f_{BB} P_{0110} (n_{A}, n_{B}) \nonumber \\
+k_{A} [(n_{A} +1) P_{0110} (n_{A}+1, n_{B}) - n_{A} P_{0110} (n_{A}, n_{B})] \nonumber \\
+k_{B} [(n_{B} +1) P_{0110} (n_{A}, n_{B}+1) - n_{B} P_{0110} (n_{A}, n_{B})] \nonumber \\
+g^A_{01} [P_{0110} (n_{A}-1, n_{B}) - P_{0110} (n_{A}, n_{B})] \nonumber \\
+g^B_{10} [P_{0110} (n_{A}, n_{B}-1) - P_{0110} (n_{A}, n_{B})]
\end{eqnarray}

\begin{eqnarray}
\frac{d P_{0010} (n_{A}, n_{B})}{dt} = \nonumber \\
+ \frac{h_{AA}}{2} [(n_{A}+2) (n_{A}+1)] P_{1010} (n_{A}+2, n_{B}) - f_{AA} P_{0010} (n_{A}, n_{B}) \nonumber \\
+ \frac{h_{AB}}{2} [(n_{B}+2) (n_{B}+1)] P_{0110} (n_{A}, n_{B}+2) - f_{AB} P_{0010} (n_{A}, n_{B}) \nonumber \\
-\frac{h_{BA}}{2} [n_{A} (n_{A}-1)] P_{0010} (n_{A}, n_{B}) + f_{BA} P_{0000} (n_{A}-2, n_{B}) \nonumber \\
+\frac{h_{BB}}{2} [(n_{B}+2) (n_{B}+1)] P_{0011} (n_{A}, n_{B}+2) - f_{BB} P_{0010} (n_{A}, n_{B}) \nonumber \\
+k_{A} [(n_{A} +1) P_{0010} (n_{A}+1, n_{B}) - n_{A} P_{0010} (n_{A}, n_{B})] \nonumber \\
+k_{B} [(n_{B} +1) P_{0010} (n_{A}, n_{B}+1) - n_{B} P_{0010} (n_{A}, n_{B})] \nonumber \\
+g^A_{00} [P_{0010} (n_{A}-1, n_{B}) - P_{0010} (n_{A}, n_{B})] \nonumber \\
+g^B_{10} [P_{0010} (n_{A}, n_{B}-1) - P_{0010} (n_{A}, n_{B})]
\end{eqnarray}

\begin{eqnarray}
\frac{d P_{1101} (n_{A}, n_{B})}{dt} = \nonumber \\
-\frac{h_{AA}}{2} [n_{A} (n_{A}-1)] P_{1101} (n_{A}, n_{B}) + f_{AA} P_{0101} (n_{A}-2, n_{B}) \nonumber \\
-\frac{h_{AB}}{2} [n_{B} (n_{B}-1)] P_{1101} (n_{A}, n_{B}) + f_{AB} P_{1001} (n_{A}, n_{B}-2) \nonumber \\
+\frac{h_{BA}}{2} [(n_{A}+2) (n_{A}+1)] P_{1111} (n_{A}+2, n_{B}) - f_{BA} P_{1101} (n_{A}, n_{B}) \nonumber \\
-\frac{h_{BB}}{2} [n_{B} (n_{B}-1)] P_{1101} (n_{A}, n_{B}) + f_{BB} P_{1100} (n_{A}, n_{B}-2) \nonumber \\
+k_{A} [(n_{A} +1) P_{1101} (n_{A}+1, n_{B}) - n_{A} P_{1101} (n_{A}, n_{B})] \nonumber \\
+k_{B} [(n_{B} +1) P_{1101} (n_{A}, n_{B}+1) - n_{B} P_{1101} (n_{A}, n_{B})] \nonumber \\
+g^A_{11} [P_{1101} (n_{A}-1, n_{B}) - P_{1101} (n_{A}, n_{B})] \nonumber \\
+g^B_{01} [P_{1101} (n_{A}, n_{B}-1) - P_{1101} (n_{A}, n_{B})]
\end{eqnarray}

\begin{eqnarray}
\frac{d P_{1001} (n_{A}, n_{B})}{dt} = \nonumber \\
- \frac{h_{AA}}{2} [n_{A} (n_{A}-1)] P_{1001} (n_{A}, n_{B}) + f_{AA} P_{0001} (n_{A}-2, n_{B}) \nonumber \\
+ \frac{h_{AB}}{2} [(n_{B}+2) (n_{B}+1)] P_{1101} (n_{A}, n_{B}+2) - f_{AB} P_{1001} (n_{A}, n_{B}) \nonumber \\
+\frac{h_{BA}}{2} [(n_{A}+2) (n_{A}+1)] P_{1011} (n_{A}+2, n_{B}) - f_{BA} P_{1001} (n_{A}, n_{B}) \nonumber \\
-\frac{h_{BB}}{2} [n_{B} (n_{B}-1)] P_{1001} (n_{A}, n_{B}) + f_{BB} P_{1000} (n_{A}, n_{B}-2) \nonumber \\
+k_{A} [(n_{A} +1) P_{1001} (n_{A}+1, n_{B}) - n_{A} P_{1001} (n_{A}, n_{B})] \nonumber \\
+k_{B} [(n_{B} +1) P_{1001} (n_{A}, n_{B}+1) - n_{B} P_{1001} (n_{A}, n_{B})] \nonumber \\
+g^A_{10} [P_{1001} (n_{A}-1, n_{B}) - P_{1001} (n_{A}, n_{B})] \nonumber \\
+g^B_{01} [P_{1001} (n_{A}, n_{B}-1) - P_{1001} (n_{A}, n_{B})]
\end{eqnarray}

\begin{eqnarray}
\frac{d P_{0101} (n_{A}, n_{B})}{dt} = \nonumber \\
+\frac{h_{AA}}{2} [(n_{A}+2) (n_{A}+1)] P_{1101} (n_{A}+2, n_{B}) - f_{AA} P_{0101} (n_{A}, n_{B}) \nonumber \\
-\frac{h_{AB}}{2} [n_{B} (n_{B}-1)] P_{0101} (n_{A}, n_{B}) + f_{AB} P_{0001} (n_{A}, n_{B}-2) \nonumber \\
+\frac{h_{BA}}{2} [(n_{A}+2) (n_{A}+1)] P_{0111} (n_{A}+2, n_{B}) - f_{BA} P_{0101} (n_{A}, n_{B}) \nonumber \\
-\frac{h_{BB}}{2} [n_{B} (n_{B}-1)] P_{0101} (n_{A}, n_{B}) + f_{BB} P_{0100} (n_{A}, n_{B}-2) \nonumber \\
+k_{A} [(n_{A} +1) P_{0101} (n_{A}+1, n_{B}) - n_{A} P_{0101} (n_{A}, n_{B})] \nonumber \\
+k_{B} [(n_{B} +1) P_{0101} (n_{A}, n_{B}+1) - n_{B} P_{0101} (n_{A}, n_{B})] \nonumber \\
+g^A_{01} [P_{0101} (n_{A}-1, n_{B}) - P_{0101} (n_{A}, n_{B})] \nonumber \\
+g^B_{01} [P_{0101} (n_{A}, n_{B}-1) - P_{0101} (n_{A}, n_{B})]
\end{eqnarray}

\begin{eqnarray}
\frac{d P_{0001} (n_{A}, n_{B})}{dt} = \nonumber \\
+ \frac{h_{AA}}{2} [(n_{A}+2) (n_{A}+1)] P_{1001} (n_{A}+2, n_{B}) - f_{AA} P_{0001} (n_{A}, n_{B}) \nonumber \\
+ \frac{h_{AB}}{2} [(n_{B}+2) (n_{B}+1)] P_{0101} (n_{A}, n_{B}+2) - f_{AB} P_{0001} (n_{A}, n_{B}) \nonumber \\
+\frac{h_{BA}}{2} [(n_{A}+2) (n_{A}+1)] P_{0011} (n_{A}+2, n_{B}) - f_{BA} P_{0001} (n_{A}, n_{B}) \nonumber \\
-\frac{h_{BB}}{2} [n_{B} (n_{B}-1)] P_{0001} (n_{A}, n_{B}) + f_{BB} P_{0000} (n_{A}, n_{B}-2) \nonumber \\
+k_{A} [(n_{A} +1) P_{0001} (n_{A}+1, n_{B}) - n_{A} P_{0001} (n_{A}, n_{B})] \nonumber \\
+k_{B} [(n_{B} +1) P_{0001} (n_{A}, n_{B}+1) - n_{B} P_{0001} (n_{A}, n_{B})] \nonumber \\
+g^A_{00} [P_{0001} (n_{A}-1, n_{B}) - P_{0001} (n_{A}, n_{B})] \nonumber \\
+g^B_{01} [P_{0001} (n_{A}, n_{B}-1) - P_{0001} (n_{A}, n_{B})]
\end{eqnarray}

\begin{eqnarray}
\frac{d P_{1100} (n_{A}, n_{B})}{dt} = \nonumber \\
-\frac{h_{AA}}{2} [n_{A} (n_{A}-1)] P_{1100} (n_{A}, n_{B}) + f_{AA} P_{0100} (n_{A}-2, n_{B}) \nonumber \\
-\frac{h_{AB}}{2} [n_{B} (n_{B}-1)] P_{1100} (n_{A}, n_{B}) + f_{AB} P_{1000} (n_{A}, n_{B}-2) \nonumber \\
+\frac{h_{BA}}{2} [(n_{A}+2) (n_{A}+1)] P_{1110} (n_{A}+2, n_{B}) - f_{BA} P_{1100} (n_{A}, n_{B}) \nonumber \\
+\frac{h_{BB}}{2} [(n_{B}+2) (n_{B}+1)] P_{1101} (n_{A}, n_{B}+2) - f_{BB} P_{1100} (n_{A}, n_{B}) \nonumber \\
+k_{A} [(n_{A} +1) P_{1100} (n_{A}+1, n_{B}) - n_{A} P_{1100} (n_{A}, n_{B})] \nonumber \\
+k_{B} [(n_{B} +1) P_{1100} (n_{A}, n_{B}+1) - n_{B} P_{1100} (n_{A}, n_{B})] \nonumber \\
+g^A_{11} [P_{1100} (n_{A}-1, n_{B}) - P_{1100} (n_{A}, n_{B})] \nonumber \\
+g^B_{00} [P_{1100} (n_{A}, n_{B}-1) - P_{1100} (n_{A}, n_{B})]
\end{eqnarray}

\begin{eqnarray}
\frac{d P_{1000} (n_{A}, n_{B})}{dt} = \nonumber \\
- \frac{h_{AA}}{2} [n_{A} (n_{A}-1)] P_{1000} (n_{A}, n_{B}) + f_{AA} P_{0000} (n_{A}-2, n_{B}) \nonumber \\
+ \frac{h_{AB}}{2} [(n_{B}+2) (n_{B}+1)] P_{1100} (n_{A}, n_{B}+2) - f_{AB} P_{1000} (n_{A}, n_{B}) \nonumber \\
+\frac{h_{BA}}{2} [(n_{A}+2) (n_{A}+1)] P_{1010} (n_{A}+2, n_{B}) - f_{BA} P_{1000} (n_{A}, n_{B}) \nonumber \\
+\frac{h_{BB}}{2} [(n_{B}+2) (n_{B}+1)] P_{1001} (n_{A}, n_{B}+2) - f_{BB} P_{1000} (n_{A}, n_{B}) \nonumber \\
+k_{A} [(n_{A} +1) P_{1000} (n_{A}+1, n_{B}) - n_{A} P_{1000} (n_{A}, n_{B})] \nonumber \\
+k_{B} [(n_{B} +1) P_{1000} (n_{A}, n_{B}+1) - n_{B} P_{1000} (n_{A}, n_{B})] \nonumber \\
+g^A_{10} [P_{1000} (n_{A}-1, n_{B}) - P_{1000} (n_{A}, n_{B})] \nonumber \\
+g^B_{00} [P_{1000} (n_{A}, n_{B}-1) - P_{1000} (n_{A}, n_{B})]
\end{eqnarray}

\begin{eqnarray}
\frac{d P_{0100} (n_{A}, n_{B})}{dt} = \nonumber \\
+\frac{h_{AA}}{2} [(n_{A}+2) (n_{A}+1)] P_{1100} (n_{A}+2, n_{B}) - f_{AA} P_{0100} (n_{A}, n_{B}) \nonumber \\
-\frac{h_{AB}}{2} [n_{B} (n_{B}-1)] P_{0100} (n_{A}, n_{B}) + f_{AB} P_{0000} (n_{A}, n_{B}-2) \nonumber \\
+\frac{h_{BA}}{2} [(n_{A}+2) (n_{A}+1)] P_{0110} (n_{A}+2, n_{B}) - f_{BA} P_{0100} (n_{A}, n_{B}) \nonumber \\
+\frac{h_{BB}}{2} [(n_{B}+2) (n_{B}+1)] P_{0101} (n_{A}, n_{B}+2) - f_{BB} P_{0100} (n_{A}, n_{B}) \nonumber \\
+k_{A} [(n_{A} +1) P_{0100} (n_{A}+1, n_{B}) - n_{A} P_{0100} (n_{A}, n_{B})] \nonumber \\
+k_{B} [(n_{B} +1) P_{0100} (n_{A}, n_{B}+1) - n_{B} P_{0100} (n_{A}, n_{B})] \nonumber \\
+g^A_{01} [P_{0100} (n_{A}-1, n_{B}) - P_{0100} (n_{A}, n_{B})] \nonumber \\
+g^B_{00} [P_{0100} (n_{A}, n_{B}-1) - P_{0100} (n_{A}, n_{B})]
\end{eqnarray}

\begin{eqnarray}\label{MS16}
\frac{d P_{0000} (n_{A}, n_{B})}{dt} = \nonumber \\
+ \frac{h_{AA}}{2} [(n_{A}+2) (n_{A}+1)] P_{1000} (n_{A}+2, n_{B}) - f_{AA} P_{0000} (n_{A}, n_{B}) \nonumber \\
+ \frac{h_{AB}}{2} [(n_{B}+2) (n_{B}+1)] P_{0100} (n_{A}, n_{B}+2) - f_{AB} P_{0000} (n_{A}, n_{B}) \nonumber \\
+\frac{h_{BA}}{2} [(n_{A}+2) (n_{A}+1)] P_{0010} (n_{A}+2, n_{B}) - f_{BA} P_{0000} (n_{A}, n_{B}) \nonumber \\
+\frac{h_{BB}}{2} [(n_{B}+2) (n_{B}+1)] P_{0001} (n_{A}, n_{B}+2) - f_{BB} P_{0000} (n_{A}, n_{B}) \nonumber \\
+k_{A} [(n_{A} +1) P_{0000} (n_{A}+1, n_{B}) - n_{A} P_{0000} (n_{A}, n_{B})] \nonumber \\
+k_{B} [(n_{B} +1) P_{0000} (n_{A}, n_{B}+1) - n_{B} P_{0000} (n_{A}, n_{B})] \nonumber \\
+g^A_{00} [P_{0000} (n_{A}-1, n_{B}) - P_{0000} (n_{A}, n_{B})] \nonumber \\
+g^B_{00} [P_{0000} (n_{A}, n_{B}-1) - P_{0000} (n_{A}, n_{B})]
\end{eqnarray}

\newpage

\newpage

\newpage

\newpage

\newpage

\newpage


\begin{thebibliography}{99}

\bibitem{Waddington}
Waddington, C. H. 1957. The Strategy of the Genes. Allen and Unwin, London, UK.


\bibitem{Wang2010}
Wang, J., L. Xu, Wang, E.K., and S. Huang. 2010. The Potential Landscape of Genetic Circuits Imposes the Arrow
of Time in Stem Cell Differentiation. {\it Biophys. J.} 99:29-39.

\bibitem{Wang2011}
Wang, J., K. Zhang, L. Xu, and E.K. Wang. 2011. Quantifying the Waddington landscape and biological
paths for development and differentiation. {\it Proc. Natl. Acad. Sci.} 108:8257-8262.




\bibitem{Elowitz}
Elowitz, M. B., and S. Leibler. 2000. A synthetic oscillatory network
of transcriptional regulators. {\it Nature} 403:335-338.








%




\bibitem{WangJCP2007}
Kim, K., D. Lepzelter, and J. Wang. 2007. Single Molecule Dynamics and Statistical Fluctuations of Gene Regulatory Networks: A Repressilator. {\it J. Chem. Phys.} 126:034702.






%





\bibitem{Arkin}
Arkin, A., J. Ross, and H.H. McAdams. 1998. Stochastic kinetic analysis of developmental pathway bifurcation in phage lambda-infected
Escherichia coli cells. {\it Genetics} 149:1633-1648.

\bibitem{Shea}
Ackers, G. K., A. D. Johnson, and M. A. Shea. 1982. Quantitative model for gene regulation by lambda phage repressor. {\it Proc. Natl. Acad. Sci.} 79:1129-1133.


\bibitem{NatureExperiment}
Austin, D. W., M. S. Allen, J. M. McCollum, R. D. Dar, J. R. Wilgus, G. S. Sayler, N. F. Samatova,
C. D. Cox, and M. L. Simpson. 2006. Gene Network Shaping of Inherent Noise Spectra. {\it Nature} 439:608-611.

\bibitem{Hornos}
Hornos, J. E. M., D. Schultz, G. C. P. Innocentini, J. Wang, A. M. Walczak, J. N. Onuchic, and P. G. Wolynes, (2005) Self-Regulating gene: An Exact Solution.
{\it Phys. Rev. E} 72:051907.

\bibitem{Walczak}
Walczak, A. M., J. N. Onuchic, and P. G. Wolynes. 2005. Absolute rate theories of epigenetic stability.
{\it Proc. Natl. Acad. Sci.} 102:18926-18931.

\bibitem{Daniel}
Schultz, D., J. N. Onuchic, and P. G. Wolynes. 2007. Understanding stochastic simulations of the smallest genetic networks. {\it J. Chem. Phys.} 126:245102.

\bibitem{Kepler}
Kepler, T. B., and T. C. Elston. 2001. Stochasticity in Transcriptional Regulation: Origins, Consequences, and Mathematical Representations. {\it Biophys. J.} 81:3116-3136.

\bibitem{Kardar}
Das, J., M. Kardar, and A.K. Chakraborty. 2007. Purely stochastic binary decisions in cell signaling models without underlying deterministic instabilities.
{\it Proc. Natl. Acad. Sci. U.S.A.} 104:18598-18963.


\bibitem{Feng2011JPCLett}
Feng, H., B. Han, and J. Wang. 2010. Dominant Kinetic Paths of Complex Systems: Gene Networks. {\it J. Phys. Chem. Lett.} 1:1836-1840.

\bibitem{Feng2011JPCB}
Feng, H., B. Han, and J. Wang. 2011. Adiabatic and Non-Adiabatic Non-Equilibrium
Stochastic Dynamics of Single Regulating Genes. {\it J. Phys. Chem. B} 115:1254-1261.

\bibitem{Feng2012BJ}
Feng, H., B. Han, and J. Wang. 2012. Landscape and Global Stability of Nonadiabatic and Adiabatic Oscillations
in a Gene Network. Accepted by {\it Biophys. J.}.

\bibitem{Weinberger}
Singh, A., and L.S. Weinberger. 2009. Stochastic gene expression as a molecular switch for viral latency. {\it Current Opinion in Microbiology} 12:460-466.

\bibitem{Xie1}
Choi, P. J., L. Cai, K. Frieda, and S. Xie. 2008. A Stochastic Single-Molecule Event Triggers Phenotype Switching of a Bacterial Cell. {\it Science} 322:442-446.

\bibitem{Arias}
Kalmar, T., C. Lim, P. Hayward, S. Munoz-Descalzo, J. Nichols, J. Garcia-Ojalvo, and A. M. Arias. Regulated Fluctuations in Nanog Expression Mediate Cell
Fate Decisions in Embryonic Stem Cells. {\it PLoS Biol.} 7:e1000149.

\bibitem{Graf}
Graf, T., and T. Enver. 2009. Forcing cells to change lineages. {\it Nature} 462:587-594.

\bibitem{Zhou}
Zhou, J.X., and S. Huang. 2010. Understanding gene circuits at cell-fate branch points for
rational cell reprogramming. {\it Trends Genet} 27:55-62.

\bibitem{Huang}
Huang, S., Y.P. Guo, G. May, and T. Enver. 2007. Bifurcation dynamics of cell fate decision in
bipotent progenitor cells. {\it Dev. Biol.} 305:695-713.

\bibitem{Hu}
Hu M., et al. 1997. Multilineage gene expression precedes commitment in the hemopoietic
system. {\it Genes Dev.} 11:774-785.

\bibitem{Gillespie}
Gillespie D. T. 1977. Exact Stochastic Simulation of Coupled Chemical Reactions. {\it J. Phys. Chem.} 81:2340-2361.

\bibitem{WangPNAS2008}
Wang, J., L. Xu, and E. K. Wang. 2008. Potential Landscape and Flux Framework of Non-Equilibrium Networks: Robustness,
Dissipation and Coherence of Biochemical Oscillations. {\it Proc. Natl. Acad. Sci.} 105:12271-12276.

\end{thebibliography}
\end{document}